\documentclass[compactaffiliation]{Interspeech}

\interspeechcameraready 

\usepackage{multirow}

\title{Evaluation of Speech Foundation Models for ASR on Child-Adult Conversations in Autism Diagnostic Sessions}

\author[affiliation={1}]{Aditya}{Ashvin}
\author[affiliation={1}, equalcontribution]{Rimita}{Lahiri}
\author[affiliation={1}, equalcontribution]{Aditya}{Kommineni}
\author[affiliation={2}]{Somer}{Bishop}
\author[affiliation={3}]{Catherine}{Lord}
\author[affiliation={1}]{~~~~~~~~~~~~~~~~~Sudarsana Reddy}{Kadiri}
\author[affiliation={1}]{Shrikanth}{Narayanan}
\vspace{-0.2cm}
\affiliation{Signal Analysis and Interpretation Laboratory}{University of Southern California}{USA}
\affiliation{Department of Psychiatry}{University of California San Francisco}{USA}
\affiliation{Semel Institute of Neuroscience and Human Behavior}{University of California Los Angeles}{USA}

\email{ashvin@usc.edu,rlahiri@usc.edu,akommine@usc.edu, Somer.Bishop@ucsf.edu, clord@mednet.ucla.edu, skadiri@usc.edu, shri@usc.edu}


\begin{document}
\maketitle

\begin{abstract}
Reliable transcription of child-adult conversations in clinical settings is crucial for diagnosing developmental disorders like Autism. Recent advances in deep learning and availability of large scale transcribed data has led to development of speech foundation models that have shown dramatic improvements in ASR performance. However, their performance on conversational child-adult interactions remains underexplored. In this work, we provide a comprehensive evaluation of ASR performance on a dataset containing child-adult interactions from autism diagnostic sessions, using Whisper, Wav2Vec2, HuBERT, and WavLM. We find that speech foundation models show a noticeable performance drop (15-20\% absolute WER) for child speech compared to adult speech in the conversational setting. Then, we fine-tune the best-performing zero-shot model (Whisper-large) using LoRA in a low-resource setting, yielding $\sim$8\% and $\sim$13\% absolute WER improvements for child and adult speech, respectively.

\end{abstract}

\section{Introduction}
\label{sec:intro}
Self-supervised learning~(SSL) models such as WavLM \cite{Chen_2022}, HuBERT \cite{hsu2021hubertselfsupervisedspeechrepresentation}, and Wav2Vec2 \cite{baevski2020wav2vec20frameworkselfsupervised} have shown significant performance gains in ASR over conventional methods through leveraging large scale unlabelled data to learn meaningful speech representations.. Recently, whisper \cite{radford2022robustspeechrecognitionlargescale} demonstrated noticeable improvements over these SSL models through supervised multitask training on an order of magnitude larger data ($\sim$ 680k hours) compared to prior works. Notably, for adult speech recognition in clean conditions, these models achieve performance comparable to humans~\cite{radford2022robustspeechrecognitionlargescale}.

Despite extensive research efforts in ASR, progress in developing robust systems capable of processing child speech especially in a conversational context has been limited. 
Also, unlike adult speech data, reliable labeled datasets for training child ASR systems are scarce and difficult to collect as well as annotate~\cite{claus2013survey}. Significant differences between adult and child speech, including pitch, linguistic and acoustic features, and pronunciation ability~\cite{lee1997analysis,lee1999acoustics} often lead to ASR systems showing subpar performance on child speech. The shorter vocal tract length and higher fundamental frequency of children's voices have also been reported as contributing factors to the increased challenges in developing ASR systems for child speech~\cite{serizel2014vocal}. 

Additionally, speech and language abnormalities stemming from neuro-developmental disorders such as Autism Spectrum Disorder (ASD) compounds the difficulty of developing child-inclusive ASR systems.

ASD encompasses a range of neuro-developmental disorders often marked by atypical social and communication patterns. 
Previous research has demonstrated that early diagnosis and intervention for ASD can have long-lasting positive outcomes~\cite{elder2017clinical}. 
During the formative years, the brain's ongoing development and plasticity increase the likelihood of effective treatment in children~\cite{camarata2014early}. 

\par 
Diagnostic sessions of ASD are administered through Autism Diagnostic Observation Schedule (ADOS) \cite{lord2012autism} protocol that is a long-form interaction session between a clinician and a child involving activities of play and conversations. Speech content from these interaction sessions provide valuable insights that characterize verbal communication characteristics of the child and in understanding the symptom severity. Developing robust ASR systems for child-inclusive interactions in the realm of ASD could help quantify some of the clinically relevant measures~\cite{lahiri2022interpersonal}. While some recent works have tried to address ASR for child speech, most of the focus is levied on child speech alone without accounting for the impact of model performance on adult speech. Additionally, most evaluations are reported under constrained conditions such as read or spontaneous speech~\cite{fan2024benchmarking,jain2023adaptation}. 

To the best of our knowledge, this is the first work attempting to evaluate state-of-the-art ASR models in the context of child-adult conversational speech. In this work, we evaluate the performance of speech foundation models on child-adult interactions from autism diagnostic sessions. 

The contributions of this work are outlined as follows:
\begin{itemize}
\item We provide a comprehensive evaluation of the performance of speech foundation models in transcribing child-adult conversational speech.
.
\item Investigated the parameter-efficient fine-tuning to improve ASR performance in child-inclusive conversational speech. This is tested in both child-only and adult-only settings.

\item Analysis of the impact of utterance length variation on WER indicates that utterances consisting of a single word or two words exhibit noticeably poorer ASR performance.

\end{itemize}

\section{Related Work}
\label{sec:background}

High-quality child speech datasets are scarce, often difficult to collect and annotate~\cite{claus2013survey}. To address this limitation, prior works in child speech recognition often rely on data augmentation methods such as perturbation~\cite{yeung2021fundamental,kathania2021using,10446889,children_gan} or voice conversion~\cite{shahnawazuddin2020voice, le2024voicebox}. In \cite{9413801}, feature normalization was proposed as an additional method to improve children’s ASR performance, alongside data augmentation. Another alternative approach involves adopting the pretraining-fine-tuning paradigm. This approach utilizes unlabelled data through self-supervised learning~\cite{fan2022towards,mohamed2022self} or annotated adult data via transfer learning~\cite{shivakumar2020transfer} in the pretraining phase. 
With the recent advancements on speech foundation models, multiple studies~\cite{kathania2022formant,fan2024benchmarking,jain2023adaptation,attia2023kid,sinha2024effect} have contributed towards benchmarking the performance of these models for different corpora containing child speech. While \cite{jain2023adaptation,attia2023kid} emphasize improving performance by adapting Whisper and its variants for various publicly available datasets containing child speech, \cite{fan2024benchmarking} investigated different fine-tuning strategies by comparing various data augmentation and parameter-efficient fine-tuning (PEFT) methods by evaluating these methods with speech foundation models on different child speech databases.

\section{Dataset}
\label{sec:typestyle}
\label{ssec:subhead}
This study uses an in-house dataset (\textit{ADOSMod3}), composed of video recordings of Autism Diagnostic Observation Schedule (ADOS) sessions. 

ADOS is a standardized diagnostic protocol to assess ASD in children. These sessions are administered by clinicians, typically are about 40-60 minutes long. They include a combination of play and interview style subtasks intended to elicit spontaneous responses to assess verbal and non-verbal communication of the child. 

The dataset contains 165 sessions, collected across two clinical sites: the University of Michigan Autism and Communication Disorders Center (UMACC) and the Cincinnati Children’s Medical Center (CCHMC). The demographic details of the children are provided in Table~\ref{tab:ADOSdataset}. 

\begin{table}[!ht]
\begin{center}
\caption{ADOSMod3 dataset demographic details.}
\label{tab:ADOSdataset}
\resizebox{0.48\textwidth}{!}{
\begin{tabular}{c c} \toprule
\textbf{Category} & \textbf{Statistics}  \\  \midrule
 Age(years) & Range: 3.58-13.17 (mean,std):(8.61,2.49)  \\  
 {\centering Age distribution} & 2.5-7.5yrs 54,7.5-10 yrs 62, $\geq$10yrs 61 \\ \midrule
 Gender & 123 male, 42 female \\ \midrule
 \\  \hline
 {\centering Number of Utterances} & 
  Child: 53001; Adult: 62110; Total: 115,111 \\ \midrule
  \multirow{4}{4em}{\centering Clinical Diagnosis} & 86 ASD, 42 Attention Deficit Hyperactivity Disorder~(ADHD)\\  
 & 14 mood/anxiety disorder\\
 & 12 language disorder\\
 & 10 intellectual disability, 1 no diagnosis\\  \bottomrule

 \end{tabular}}
\end{center}
\vspace{-0.4cm}
\end{table}
\par 
Ground truth speech is annotated for Emotions, Social difficulties and annoyances subtasks, which are chosen for their ability to elicit free-form responses from the child. Thereby, maintaining a balanced distribution of speech between the child (46\%) and the adult (54\%) as shown in Table~\ref{tab:ADOSdataset}, unlike play tasks that may predominantly feature clinician speech. In total, the dataset comprises 8 hours of child speech and 9 hours of adult speech. Each subtask is approximately three minutes long. The dataset was manually transcribed following SALT guidelines, annotating for speaker identity (adult/child), time boundaries, and associated utterances. 

During preprocessing, tags related to laughter, breathing, and inaudible speech were excluded, as conventional ASR systems are unable to accurately capture these elements. 
 
Child utterances are generally shorter and more variable in length compared to adult speech, reflecting the spontaneous and often fragmented nature of conversations between children and adults. This variability impacts ASR performance and underscores the necessity for targeted fine-tuning to address the challenges associated with conversational speech.

\section{Experiments}
\begin{figure*}[ht]
\centering
  \includegraphics[width=\textwidth]{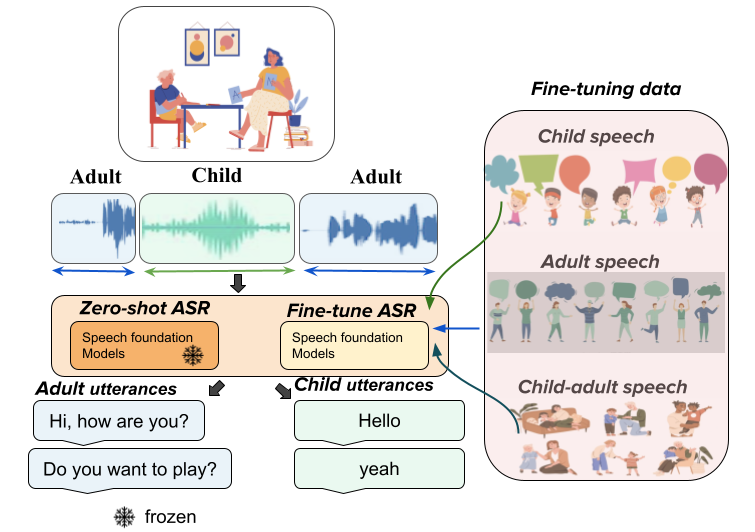}
  \caption{Schematic diagram of the experimental pipelines used in this study, showcasing the zero-shot and fine-tuned ASR systems applied to child-adult conversations.}
  \label{fig:flowchart}
\end{figure*}

Figure~\ref{fig:flowchart} provides a schematic overview of the experimental pipelines used in this study to evaluate the performance of zero-shot and fine-tuned ASR systems on child-adult conversational data. Both approaches rely on pre-trained speech foundational models for ASR.
\subsection{Zero-shot experiments}
Zero-shot evaluation of ASR is performed across aforementioned state-of-the-art speech foundation models. The details of the models are provided in Table~\ref{table:pretrained_models}. We include evaluation of model trained on supervised learning objectives, such as Whisper, and using self-supervised objectives, such as Wav2Vec2, HuBERT, and WavLM. For all the self-supervised models, the corresponding large variants of the models are used. In the case of Wav2Vec2, an additional model employing self-training~\cite{xu2020selftrainingpretrainingcomplementaryspeech} is also evaluated. For the Whisper, we assess the performance of base, medium, and large variants to investigate the impact of parameter scaling on ASR performance. WER for all the experiments is reported on the entire dataset of $\sim$17hours.

\begin{table}[]
\caption{Summary of the state-of-the-art speech foundation models used in this study.}
    \footnotesize
    \begin{tabular*}{\linewidth}{lcccc}
        \toprule
        
        \multirow{2}{*}{\shortstack{\textbf{Pre-trained}\\\textbf{Architecture}}} & 
        \multirow{2}{*}{\textbf{Input}} & 
        \multirow{2}{*}{\shortstack{\textbf{Layers}}} &
        \multirow{2}{*}{\shortstack{\textbf{Hidden}\\\textbf{Size}}} & 
        \multirow{2}{*}{\shortstack{\textbf{Params}}}  \\ 

        & & & & \\ 
         
        \midrule
        \textbf{Whisper base} & Mel-Spec & 6 & 512 & 72M \\ 
        \textbf{Whisper medium} & Mel-Spec & 12 & 768 & 244M \\
        \textbf{Whisper large-v3} & Mel-Spec & 32 & 1280 & 1.55B \\
        \textbf{Wav2Vec 2.0} & Raw Audio & 24 & 1024 & 317M \\ 
        \textbf{WavLM large} & Raw Audio & 24 & 1024 & 343M \\ 
        \textbf{HuBERT large} & Raw Audio & 24 & 1024 & 317M \\   
        \bottomrule
    \end{tabular*}
\vspace{-2.5mm}
\label{table:pretrained_models}
\end{table}

\subsection{Fine-tuning experiments}
Following zero-shot evaluation, the best-performing model is selected for fine-tuning. The dataset is partitioned into training and testing sets using a 70/30 split ($\sim$12 hours for training and $\sim$5 hours for testing), ensuring a between-subjects division (all utterances from a given child are either in the training or testing set). We consider two fine-tuning scenarios: adult speech only and child speech only. The adult-only and child-only settings restrict the training set of fine-tuning to utterances of adults and children respectively. 

Parameter Efficient fine-tuning method, Low Rank Adaptation (LoRA)~\cite{hu2021loralowrankadaptationlarge} with rank set to 32 is used for fine-tuning whisper large-v3 model since these techniques have been shown to provide better performance improvements in low resource scenarios. 

We use Adam optimizer with a learning rate of 2e-5, 1000 warm-up steps, and a cosine learning rate scheduler. Fine-tuning is done on a single A40 GPU with a batch size of 8 and gradient accumulation step of 2. Word Error Rate (WER) is reported for both zero-shot and fine-tuning experiments. From session level audio, child and adult utterances are derived from the annotations and transcripts for each of these utterances are generated. The fine-tuning settings i.e., child only and adult only can be correspondingly generated from these utterances. 

\section{Results}
\label{sec:print}
\subsection{Zero-shot Evaluation Results}
Table~\ref{table:wer-zeroshot} shows the zero-shot ASR performance of speech foundation models. Results are reported for child speech, adult speech, and combined speech (both adult and child) on the entire dataset. Whisper family models perform noticeably better than the other self-supervised models. This could be owing to whisper being trained on data recorded in multiple environments, recording conditions and with larger speaker diversity \cite{radford2022robustspeechrecognitionlargescale}. Within whisper models, large attains a WER of 55.78\% for child speech, 25.96\% for adult speech, and 38.66\% for the combined dataset, significantly outperforming both the base and medium versions. This trend suggests that increasing model size correlates with improved performance, likely due to enhanced capacity to generalize from diverse speech data. Hence, all the fine-tuning experiments are preformed using whisper-large. 
\par 
A significant performance gap between child speech and adult speech WER is observed across all models. This is indicative of the scarcity in child speech during training of speech foundation models. 

\begin{table}[h]
\centering
\caption{Zero-shot WER comparison for state of the art speech foundation models. Combined refers to the setting with both adult and child speech. Lower WER is better.}
\vspace{0.1cm}

\label{table:wer-zeroshot}
\begin{tabular}{lccc}
\toprule
\centering\textbf{Model} & \centering\textbf{Child} & \centering\textbf{Adult} & \textbf{Combined} \\ \midrule
Whisper \\
\quad base & \centering 77.15 & \centering 39.85 & 55.48\\ 
\quad medium & \centering 58.51 &  27.77 & 40.68\\ 
\quad large-v3 & \centering \textbf{55.78} & \textbf{25.96} & \textbf{38.66}\\ \midrule
HuBERT   & \centering 83.45 &  51.96 & 64.39\\ \midrule
Wav2Vec2\\
\quad large & \centering 92.63 &  67.01 & 77.41\\ 
\quad self training & \centering 82.35 &  52.68 & 64.38\\ \midrule
WavLM & \centering  89.31 &  61.8 & 72.97\\ \bottomrule
\end{tabular}
\end{table}
\begin{table}[]
\centering
\caption{WER for fine-tuned Whisper Large models. Lower WER is better. Child only and adult only refer to the training set during fine-tuning.}
\vspace{0.1cm}
\label{table:fine-tuning}
\resizebox{8.2cm}{1.2cm}{
\begin{tabular}{lccc}
\toprule
\centering\textbf{Setting} & \centering\textbf{Child WER} & \centering\textbf{Adult WER} & \textbf{Combined WER} \\ \midrule
Zero-shot   & \centering 54.04 & \centering 26.53 & 37.59 \\ \midrule
\textit{Fine-tuning} \\
\quad Adult only & \centering  \textbf{46.01} &  \textbf{13.09} & {29.29}\\ 
\quad Child only & \centering  46.26 &  13.2 & \textbf{28.99}\\ \bottomrule
\end{tabular}}
\end{table}

\subsection{Fine-tuning Evaluation Results}
\label{ssec:subhead}
During fine-tuning, for settings (child only and adult only), the testing data remains the same. The fine-tuning results are shown in Table~\ref{table:fine-tuning}, demonstrate improvement in WER compared to zero-shot performance, for both child only and adult only fine-tuning settings. Fine-tuning on either child speech only or adult speech only, resulted in reduction of $\sim$8\% WER for children speech and $\sim$13\% WER for adult speech on test set compared to zero-shot performance.

In all fine-tuning settings, the relative improvement in adult speech WER is higher than child speech. Interestingly, fine-tuning exclusively on either child speech only or adult speech only gave higher improvements in WER for both child and adult, indicating the model could potentially be learning the characteristics of the channel during fine-tuning thereby leading to a significant improvement. 

\subsection{Impact of Utterance Length on WER}
\label{ssec:subhead}
\begin{figure}[ht]
  \includegraphics[scale=0.098]{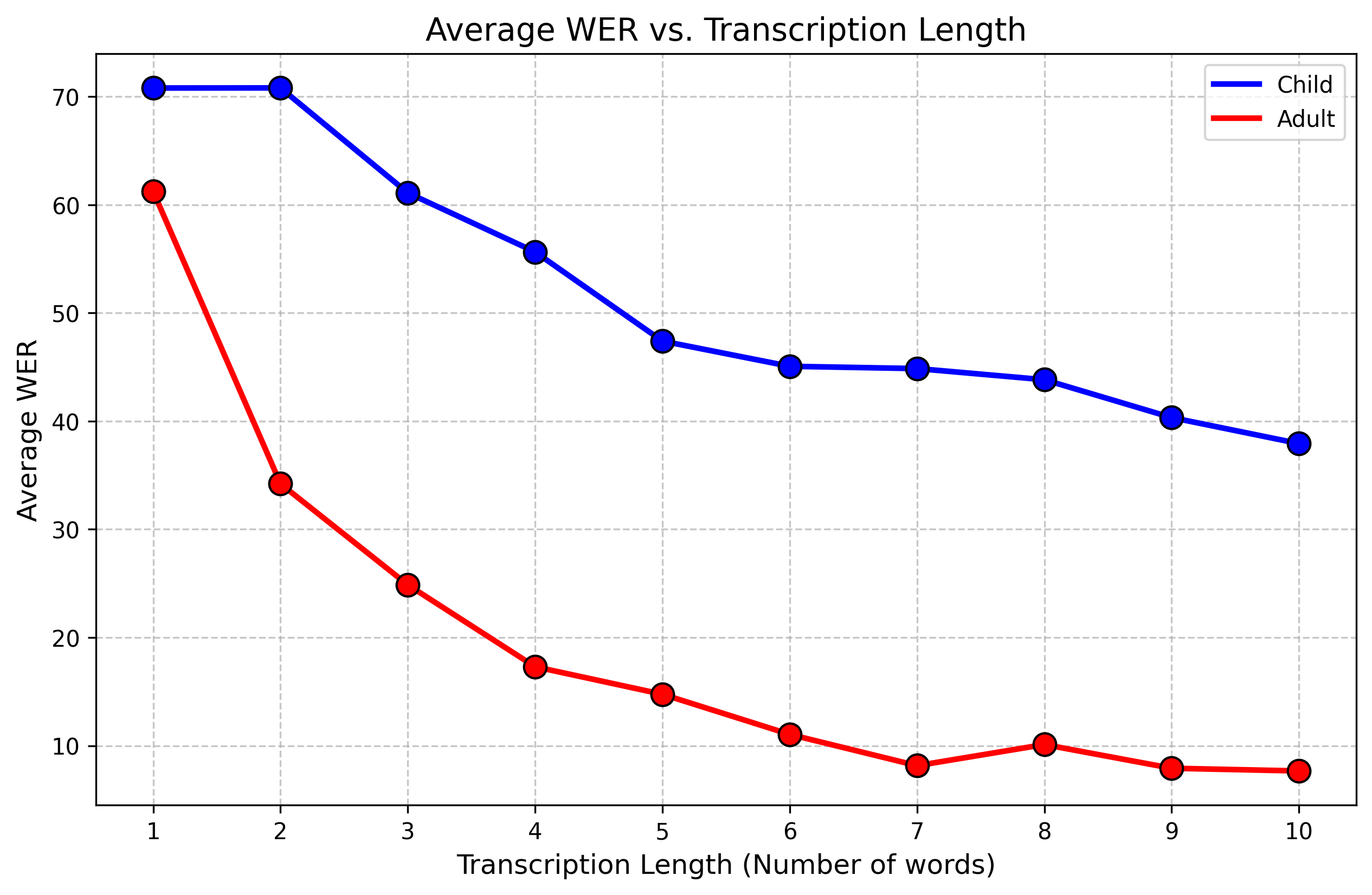}
  \vspace{-0.3cm}
  \caption{Plot of average WER with number of words in an utterance.}
  \label{fig:wer-vs-length}
\end{figure}

To better comprehend the performance of whisper large-v3 model, we analyse the relationship between transcription length (in number of words) and WER in Figure \ref{fig:wer-vs-length}. 
As the number of words within an utterance increases, a decrease in WER is observed, with single-word utterances having the highest WER.

\begin{figure}[ht]
  \includegraphics[scale=0.098]{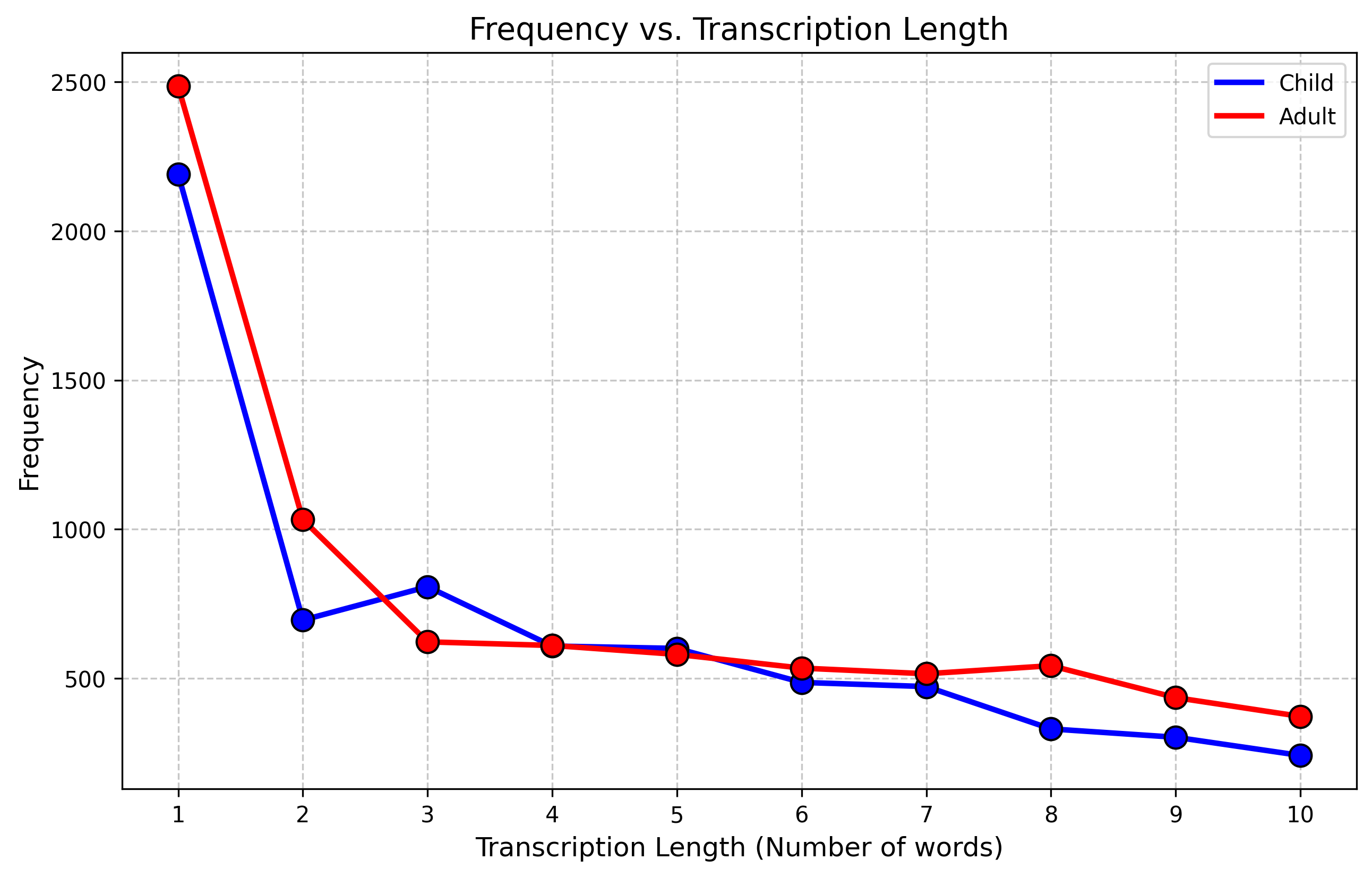}
    \vspace{-0.5cm}
  \caption{Frequency of words with increasing transcription length.}
  \label{fig:freq-vs-length}
\end{figure}

Figure \ref{fig:freq-vs-length} shows that the frequency of transcripts decreases as the number of words in said transcript increase. This shows that the large number of single-word and two-word transcriptions plays a substantial role in the higher WER.
The findings suggest that shorter transcriptions, particularly those comprising one or two words, are more prone to identification errors, resulting in higher WER. The probable reason is that the Whisper Large model is trained on audio chunks of 30 seconds each. 
\section{Conclusion}
\label{sec:page}
In this study, we investigated the performance of speech foundation models for ASR on conversational speech involving children (with autism) and adults. Our experiments focused on both zero-shot evaluation and fine-tuning to understand the benefits of adaptation of such models to adult-child dyadic interaction domain. Whisper Large V3 model emerged as the most efficient model consistently yielding superior performance as compared to the other candidates across all the experiments including adult speech, child as well as combined child-adult speech. In fact the performance of the Whisper model further improved with fine-tuning resulting in 8\% of WER for child speech and 13\% of WER for adult speech, compared to zero-short evaluation. In the future, we plan to extend our evaluation to other datasets that include child speech and perform cross-dataset evaluations to assess generalizability. Furthermore, we aim to experiment with various data augmentation strategies within the context of child-inclusive dyadic conversational speech to enhance ASR performance.

\section{Acknowledgements} We acknowledge \textit{Simons Foundation} for the funding.

\bibliographystyle{IEEEbib}
\bibliography{main}

\end{document}